\documentclass[aps,prd,twocolumn,superscriptaddress,nofootinbib,preprintnumbers]{revtex4-2}
\usepackage{amsmath}
\usepackage{graphicx}
\usepackage{subfigure}
\usepackage{amssymb}
\usepackage{xcolor}
\usepackage{multirow}
\usepackage{cancel}
\usepackage{color}
\usepackage{ulem}
\usepackage{listings}
\usepackage{xcolor}
\usepackage{float}
\usepackage{slashed}
\usepackage{wrapfig}
\usepackage{mathtools}
\usepackage[colorlinks,citecolor=blue]{hyperref}
\usepackage[T1]{fontenc}  
\usepackage[utf8]{inputenc}  
\usepackage{times}
\begin{document}
	
\title{Particle decay as asymptotic  narrow parametric resonance}

\author{Shao-Ping Li}
\email{lisp@het.phys.sci.osaka-u.ac.jp}
\affiliation{Department of Physics, The University of Osaka, Toyonaka, Osaka 560-0043, Japan}

\begin{abstract}
Parametric resonance can produce particles from oscillating scalar field, where an exponential growth of the particle number density can be developed. While it has been noticed that stimulated decay in Boltzmann equations exhibits similar parametric dependence of the exponential growth, it is not  quantitatively clear yet under what circumstance can the two phenomena reconcile. We demonstrate that trilinear particle interaction   in the Boltzmann equation can   provide a good approximation to describe the distribution function and the number density in the asymptotic regime of  narrow parametric resonance.  We find that the crucial treatment leading to the quantitative agreement  is a proper  Gaussian simulation of  the momentum spread inherited from nonrelativistic particle decay. With the simple particle picture, the analytic Boltzmann approximation can be applied to explosive photon production from axions/axion-like particles and to  dark matter production from oscillating fields.   
\end{abstract}
	
\maketitle

\preprint{OU-HET 1275} 
	
\section{Introduction}	\label{sec:intro}
It has been known for decades that particles can be produced abundantly in a background of  oscillating scalar/pseudoscalar  fields, previously introduced as a short period of preheating  right after inflation but prior to reheating~\cite{Kofman:1994rk,Kofman:1997yn}. Over the years, this production mechanism has been widely considered as an efficient particle   process for  ultralight dark matter production~\cite{Agrawal:2017eqm,Co:2017mop,Dror:2018pdh}, photon production from axion clump condensates~\cite{Tkachev:2014dpa,Hertzberg:2018zte,Arza:2018dcy}, primordial black hole  formation~\cite{Suyama:2006sr,Cai:2018tuh}, gravitational wave generation~\cite{Kawai:2017kqt,Cai:2020ovp}, axion production and  fragmentation~\cite{Fonseca:2019ypl,Co:2020dya}, among others.

Parametric resonance may be understood as an induced process~\cite{Kasuya:1996np}, where  particles produced at late times depend on those at early times. This pattern was qualitatively  confirmed in the so-called narrow parametric resonance (NPR)~\cite{Kofman:1994rk,Shtanov:1994ce,Yoshimura:1995gc,Kofman:1997yn}, where the initial amplitude of the oscillating field is small. In the NPR mechanism, a small initial amplitude allows a perturbative treatment to evaluate the particle growth rate either using the   Floquet theorem~\cite{Amin:2014eta} or   the Mathieu equation~\cite{Kovacic2018}. The growth rate can feature an exponential behavior in time, $\Gamma\propto e^{q m t}$, where $m$ denotes the mass of the oscillating field and $q\ll1$  corresponds to the typical NPR regime.

In recent years, there is an increasing attention on the equivalence between  NPR and  stimulated decay based on the Boltzmann equation. Typically, NPR is formulated by equations of motion derived in classical field theory, while the semi-classical Boltzmann equations feature  particle collisions derived in the $S$-matrix formalism of quantum field theory (QFT). Since oscillating field condensates    reflect a  collection of numerous nonrelativistic particles, it was argued previously that these condensates can be formulated in QFT, leading to a decay rate   equivalent to  the $S$-matrix formalism~\cite{Matsumoto:2007rd}, which was   confirmed earlier by using a Schr\"{o}dinger   picture of QFT~\cite{Yoshimura:1995gc,Yoshimura:1996fk}.  For bosonic particle production, Boltzmann equations   exhibit an exponential growth due to final-state Bose enhancement. This was noticed in photon production from axion/axion-like fields using the uncertainty principle~\cite{Alonso-Alvarez:2019ssa}, however, the quantitative dependence of the growth rate, by default, cannot be precisely determined.

 Indeed, if one  follows the Boltzmann equation of axion decay to photons $a\to 2\gamma$, the stimulated decay rate reads
\begin{align}
\Gamma_{\rm stim}=\Gamma_{0} (1+2f_\gamma)\,,
\end{align}
where $f_\gamma$ is the final-state photon distribution function from axion production, with the spontaneous decay rate $\Gamma_{0}$. 
From the photon number  Boltzmann equation  
\begin{align}
\frac{dn_\gamma}{dt}=2\Gamma_{\rm stim} n_a\,,
\end{align}
one can derive  an exponential solution in the limit $f_\gamma\gg 1$ 
\begin{align}
	n_\gamma\propto e^{\tilde\mu t}\,,
\end{align}
by substituting the relation $f_\gamma=n_\gamma/(2\pi^2 k^2 \delta k)$. 
The exponential growth rate $\tilde{\mu}$ depends crucially on the momentum spread $\delta k$ that is determined by nonrelativistic axion decay. Following this approach,  the growth rate was estimated in Ref.~\cite{Carenza:2019vzg} to be  $\tilde{\mu}=qm_a\pi/4$, which differs from $\tilde{\mu}=q m_a/2$ predicted by    NPR~\cite{Kofman:1997yn}. While the numerical  difference in the exponent is small, the difference will become increasingly large after several periods of axion oscillations, namely when $m_at\gg 1$.  It is  also recently shown that starting from a distribution function constructed  in QFT, the correct exponential growth rate can be matched to NPR~\cite{Moroi:2020bkq}. However, this QFT-constructed distribution function was later inserted into  the Boltzmann equation to calculate the number density,  leading to a conclusion that the Boltzmann equation can yield a number density orders of magnitude larger than   NPR in the  asymptotic limit $m_at\gg 1$.

In this work, we demonstrate that the distribution function directly obtained from the Boltzmann equation can provide a   good   approximation to describe the exponential growth in the asymptotic NPR regime.   The resulting number density can also  approximately work to predict the explosively increasing  number density before   backreaction comes into play. It provides a simple analytic approach to determine the final particle number density that can have   impacts on observables in the early Universe. In particular, it allows one to analyze  late-time thermalization processes between  produced particles and the background plasma, where   Boltzmann equations have been well established e.g., in  the  big-bang nucleosynthesis (BBN) and cosmic microwave background (CMB).  

We begin in section~\ref{sec:Mathieu} to calculate the distribution function, or the  occupation number,  within the field theory, by taking the   trilinear interaction as a prototype scenario. Then we show in section~\ref{sec:Boltzmann} how the evolution of the distribution function obtained from the Boltzmann equation can be matched to that predicted in the asymptotic NPR regime. We discuss in section~\ref{sec:disc}  the application of the simple Boltzmann equation and its validity in particle production from oscillating fields, especially in  regimes where cosmic  expansion, backreaction and scattering with a background thermal plasma can be neglected.  Finally,  we draw the conclusion  in section~\ref{sec:con}.

\section{Trilinear parametric resonance}\label{sec:Mathieu}
Let us consider the typical Yukawa or trilinear  interaction 
\begin{align}\label{eq:lag}
	\mathcal{L}=-\frac{1}{2}\mu \phi \chi^2\,,
\end{align}
where $\phi$ denotes the  background oscillating scalar  field and $\chi$ a bosonic particle.  We parameterize the $\phi$ evolution after the oscillation begins at some initial time $t_i$, with the field amplitude
\begin{align}\label{eq:cos-phi}
	\phi=\phi_i \left(\frac{a(t_i)}{a(t)}\right)^{3/2}\cos(m_\phi t)\equiv \bar \phi \cos(m_\phi t)\,,
\end{align}
where $a(t)$ is the scale factor as a function of the cosmic time $t$ and $m_\phi$ the mass of $\phi$. For definiteness, we focus on the regime where $m_\chi\ll m_\phi$ and hence neglect  the mass of $\chi$. 
We assume that right after the oscillation begins, the kinetic energy of $\phi$ is negligible such that the energy density of nonrelativistic $\phi$ gives
\begin{align}
 \rho_\phi\approx \frac{1}{2}m_\phi^2 \bar \phi^2\,, 
\end{align}
and the number density  scales like nonrelativistic matter in the radiation-dominated  Universe,
\begin{align}\label{eq:nphi}
	n_\phi=\frac{1}{2}m_\phi \phi_i^2 \left(\frac{T}{T_i}\right)^{3}\propto a(t)^{-3}\,,
\end{align}
with $T$ the cosmic temperature. 

In  Fourier space, the   mode function of $\chi$, denoted as  $u_k(t)$ with the  wave number $k\equiv |\vec  k|$ and a mass dimension of $-1/2$,  obeys the Klein-Gordon equation
\begin{align}
\frac{d^2 u_k}{dt^2}+\omega_k^2(t) u_k=0\,,
\end{align} 
with the time varying frequency $\omega_k(t)=\sqrt{k^2-\mu \phi(t)}$. 
If we can neglect the Hubble expansion, backreaction to $\phi$ from accumulated $\chi$, and  any interaction between the produced $\chi$ with a background thermal plasma, the above equation can be converted into the Mathieu equation~\cite{Kovacic2018}
\begin{align}\label{eq:Mathieu}
	\frac{d^2 u_k}{d z^2}+(A_k-2q\cos(2z)) u_k=0\,,
\end{align}
where 
\begin{align}
	A_k\equiv \frac{4k^2}{m_\phi^2}\,, \quad q\equiv \frac{2\mu \bar \phi}{m_\phi^2}\,, \quad z\equiv \frac{m_\phi t}{2}\,.
\end{align}
The parameter $q$  determines whether the resonance is in the narrow ($q\ll 1$) or broad regime ($q\gg 1$)~\cite{Fujisaki:1995dy,Fujisaki:1995ua,Khlebnikov:1996wr,Kofman:1997yn,Dufaux:2006ee}.  When $A_k-2q<0$, $\chi$ carries a negative effective mass. This regime   can lead to  tachyonic instability, though it generally occurs for a large $q$~\cite{Dufaux:2006ee}.  The Mathieu equation can be   solved numerically with the following  initial conditions
\begin{align}
	u_k(0)&=\frac{1}{\sqrt{2k}}\,,
\\[0.2cm]
	\frac{d u_k}{dt}\Big|_{t=0}&=\frac{m_\phi}{2}\frac{du_k}{dz}\Big|_{z=0}=-i \sqrt{\frac{k}{2}}\,,
\end{align}
where we have set $t_i=0$ for simplicity. Under this setup, the above   conditions correspond to a plane-wave state, where no $\chi$ particle is produced initially. For clarity, we will set $m_\phi=1$~eV throughout, but the conclusions drawn below do not depend on this scale choice. 

The occupation number  is constructed by the mode function via~\cite{Kofman:1997yn}
	\begin{align}\label{eq:n_k-def}
	n_k&=k\left(\frac{|d u_k/dt|^2}{k^2}+|u_k|^2\right)-1
	\\[0.2cm]
	&=\frac{m_\phi^2}{4k}\left|\frac{d u_k}{dz}\right|^2+k|u_k|^2-1\,,
\end{align} 
where we have multiplied the occupation number by a factor of 2 to account for the pair production of $\chi$. This is more clear when $\chi$ is identified as photons, where   two helicity states are produced~\cite{Agrawal:2017eqm}.
The particle number and energy densities are obtained by integrating the occupation number in the whole momentum space, 
\begin{align}\label{eq:nt-def}
n_\chi^{\rm M}(z)&=\int \frac{d^3k}{(2\pi)^3} n_k(z)\,,
\\[0.2cm]
\rho_\chi^{\rm M}(z)&=\int \frac{d^3k}{(2\pi)^3} k\, n_k(z)\,.\label{eq:rhot-def}
\end{align}
with the superscript M for results obtained from the Mathieu equation.

It is worthwhile to mention that the trilinear interaction can also be specified as   photon production from ultralight dark matter, axions or axion-like particles, where  the generic Lagrangian reads
\begin{align}\label{eq:lag2}
	\mathcal{L}=-\frac{g_{\phi\gamma}}{4}\phi F_{\mu\nu} \tilde{F}^{\mu\nu}\,,
\end{align}
with $F_{\mu\nu}$ the  photon field strength tensor and its dual $\tilde{F}_{\mu\nu}$. The resulting Mathieu equations read
\begin{align}\label{eq:Mathieu2}
	\frac{d^2 u^\pm_k}{d z^2}+(A_k\mp 2q\cos(2z)) u^\pm_k=0\,,
\end{align}
for two helicity states of photons $u^\pm_k$, with 
\begin{align}
A_k=\frac{4k^2}{m_\phi^2}\,,\quad q\equiv \frac{2g_{\phi\gamma} \bar \phi k}{m_\phi}\,,\quad z\equiv \frac{m_\phi t}{2}-\frac{\pi}{4}\,.
\end{align}
The number and energy  densities in this case sum over the two helicity states $u_k^\pm$~\cite{Agrawal:2017eqm}.

Note that the $\chi$-particle production can also be formulated in terms of the Bogoliubov approach, as widely applied to particle creation in curved spacetime~\cite{Parker:1969au}. 
Potential  relations between the Bogoliubov and Boltzmann approaches were discussed in different regimes~\cite{Asaka:2010kv,Kaneta:2022gug}. Here, we will not delve  into the identification of potential equivalence between the two approaches, but instead focus on how $n_k$ and $n_\chi^{\rm M}$ numerically obtained from Eq.~\eqref{eq:Mathieu} can be well described by the Boltzmann equation. The calculation performed below can be straightforwardly applied to  Eq.~\eqref{eq:lag2}, though it will not be detailed in this work. 
	 
\section{Gaussian Boltzmann equation}	\label{sec:Boltzmann}
When neglecting Hubble expansion and scattering with background particles, we can write the Boltzmann equation for distribution function $f_\chi$ that is determined by a collection of  $\phi$  particles depleting  into $\chi$ particles, $j \phi\to 2\chi$
\begin{align}\label{eq:dfdt}
	\frac{\partial f_{\chi}}{\partial t}&=\sum_{j=1}^{\infty}\frac{1}{k_j}\int dP_\phi dP'_\chi (2\pi)^4 \delta^{4}(k)	\left|\mathcal{M}_j\right|^2\mathcal{F}\,,
\end{align}
where we have multiplied the right-hand side by a factor of 2 to account for pair production of $\chi$. In Eq.~\eqref{eq:dfdt},  the phase-space differential is defined by
\begin{align}
dP_\phi\equiv	\prod_{i=1}^{j} \frac{d^3 k_i}{(2\pi)^2 2E_i}\,,\quad dP'_\chi\equiv
\frac{d^3 k'_{\chi}}{(2\pi)^3 2E'_{\chi}}\,,
\end{align}
and the statistics function reads
\begin{align}
	\mathcal{F}&\equiv \left(\prod_{i=1}^{j}f_i\right)(1+f_\chi)(1+f'_\chi)-f_
	\chi f'_\chi \prod_{i=1}^{j}(1+f_i)
	\\[0.2cm]
	&\approx \left(\prod_{i=1}^{j}f_i\right) \left(1+2f_\chi\right).
\end{align}
where   $f_i$ is  the distribution function of $\phi$, and the second approximation holds in the limit $f_i\gg 1$ and $f_{\chi}\approx f'_\chi$.

The general formula for the   amplitude of $j\phi\to 2\chi$ can be obtained from  Feynman diagrams. In the limit of $p_i=0$, we arrive at 
\begin{align}
	\mathcal{M}_j=j! \mu^j \prod_{i=1}^{j-1} \frac{1}{(i^2-i j)m_\phi^2}=
\frac{j! (-1)^{j-1}}{[(j-1)!]^2}\frac{\mu^j}{m_\phi^{2(j-1)}}\,,
\end{align}
where the factor of $j!$ appears due to the degeneracy of Feynman diagrams, namely the number of Feynman diagrams  due to  permutation of  momenta~\cite{Li:2024bbe}.
The squared amplitude after taking into account a symmetry factor of $1/(2j!)$ yields
\begin{align}
	|\mathcal{M}_j|^2=\frac{j! }{2[(j-1)!]^4}\frac{\mu^{2j}}{m_\phi^{4(j-1)}}\,.
\end{align}
We have checked  this result by building a model file in \texttt{FeynArts}~\cite{Hahn:2000kx} and calculate the amplitude  in \texttt{FeynCalc}~\cite{Shtabovenko:2023idz}.

Integrating $dP'_\chi$ via $\delta^3(\sum \vec k_i-\vec k_\chi -\vec k'_\chi)$ in Eq.~\eqref{eq:dfdt} gives rise to 
\begin{align}\label{eq:dfdt2}
	\frac{\partial f_\chi}{\partial t}&=\sum_{j=1}^{\infty}\frac{\pi |\mathcal{M}_j|^2}{k^2_j}\left(\frac{\bar\phi}{2}\right)^{2j} \delta(j m_\phi-2k_j)(1+2f_\chi)
	\\[0.2cm]
	&\approx \frac{\pi m_\phi^4}{2k_1^2}\left(\frac{q}{4}\right)^2 \delta(m_\phi-2k_1)(1+2f_\chi)\,,\label{eq:dfdt3}
\end{align}
where we have used the approximation of nonrelativistic $\phi$ 
\begin{align}
\int \prod_{i=1}^{j} \frac{d^3 k_i}{(2\pi)^2 2E_i}f_{i}=\left(\frac{n_\phi}{2m_\phi}\right)^j=\left(\frac{\bar \phi}{2}\right)^{2j},
\end{align}
with Eq.~\eqref{eq:nphi} used.  The result given by   Eq.~\eqref{eq:dfdt3} is obtained by taking only the decay channel $j=1$ with $k_1=m_\phi/2$.  This approximation results from the fact that the product of $|\mathcal{M}_j|^2$ and $\bar \phi^{2j}$ yields a factor of $(q/4)^{2j}$, which is dominated by the $j=1$ channel for  $q\ll 1$. This coincides with the common lore in NPR that the resonance production dominantly occurs in the first bandwidth where $A_k=1$.

\begin{figure*}
	\centering
	\includegraphics[scale=0.56]{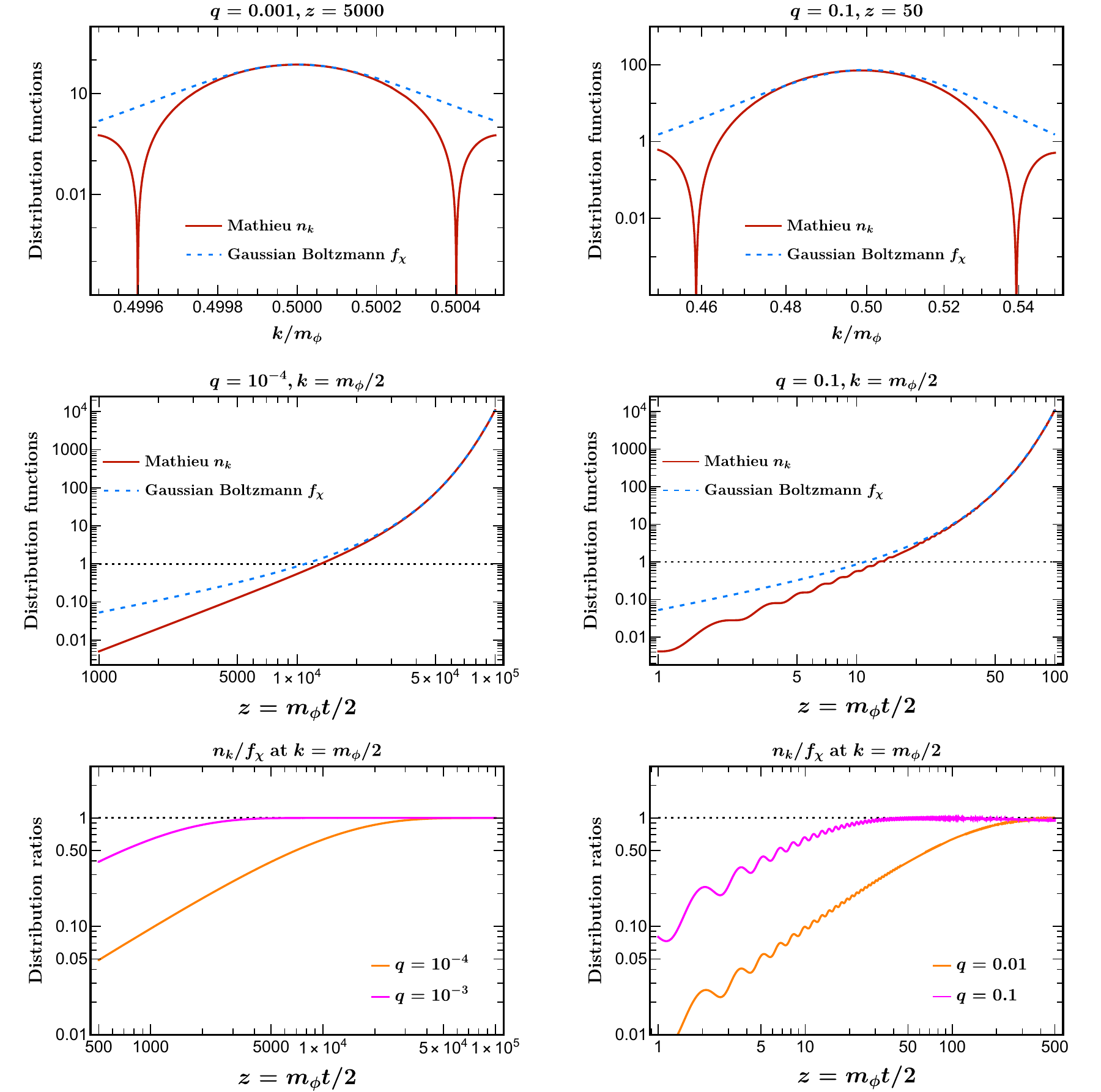} 
	\caption{\label{fig:distribution} \textit{Top}: the peak structures of the distribution function $n_k$ obtained from the  Mathieu equation, Eq.~\eqref{eq:n_k-def}, and $f_\chi$ from the Boltzmann equation, Eq.~\eqref{eq:fchi-z}. \textit{Middle}: the time evolution of distribution functions at the resonance peak $k=m_\phi/2$. \textit{Bottom}: the time evolution of the ratio $n_k/f_\chi$ at $k=m_\phi/2$.
	}
\end{figure*}

To proceed with Eq.~\eqref{eq:dfdt3}, we simulate the Dirac $\delta$-function with a Gaussian distribution
\begin{align}\label{eq:Gaussian}
	\delta(m_\phi- 2k_1)\approx \frac{a}{\sigma }e^{-\frac{(k_1- m_\phi/2)^2}{b\sigma^2}}\,,
\end{align}  
where the simulation coefficients $a, b$ are determined by the normalization condition 
\begin{align}
	\int_0^\infty  \frac{a}{\sigma}e^{-\frac{(k_1- m_\phi/2)^2}{b\sigma^2}}dk_1=\frac{1}{2}\,,
\end{align}
leading to 
\begin{align}
 a\sqrt{b}=\frac{1}{2\sqrt{\pi}}\,.
\end{align}
The width $\sigma$ describes the momentum spread inherited from nonrelativistic $\phi$ decay, which can be perturbatively calculated from the effective frequency with $k^2\gg \mu\phi(t)$. At leading order, it reads
\begin{align}\label{eq:LOwidth}
 k_1\left(1-\frac{\mu \bar\phi}{2k_1^2}\right)\leqslant \omega_k\leqslant k_1\left(1+\frac{\mu \bar\phi}{2k_1^2}\right).
\end{align}
Therefore, the momentum has a spread 
\begin{align}
	\sigma\equiv \delta k_1=\frac{\mu\bar\phi}{k_1}=q m_\phi\,,
\end{align}
which is the same as that derived from Eq.~\eqref{eq:Mathieu} by using e.g., the \textit{harmonic balancing}~\cite{Kovacic2018}.  Substituting Eq.~\eqref{eq:Gaussian} into Eq.~\eqref{eq:dfdt3} with    the  following simulation  coefficients 
\begin{align}\label{eq:simulation-1}
	a=\frac{2}{\pi}\,, \quad b=\frac{\pi}{16}\,,
\end{align}
we arrive at 
\begin{align}\label{eq:fchi-z}
	f_\chi(z)=\frac{1}{2}e^{\mathcal{G}qz}-\frac{1}{2}\,,
\end{align}
which has the same growth rate $qm_\phi/2$ as in the NPR regime except for the Gaussian modification
\begin{align}\label{eq:G}
	\mathcal{G}\equiv \text{exp}\left(-\frac{16(k_1-m_\phi/2)^2}{\pi q^2 m_\phi^2}\right)\,,
\end{align}
having the peak at $k_1=m_\phi/2$.

The resulting number density of $\chi$ in the Boltzmann approach is then  obtained by integrating Eq.~\eqref{eq:fchi-z} over   momentum space, 
\begin{align}\label{eq:nchi-B}
	n_{\chi}^{\rm B}(z)&=\int \frac{d^3 k}{(2\pi)^2}f_{\chi}(z)
	\\[0.2cm]
	&\approx \frac{1}{2\pi^2} \int_{k_1-n \sigma}^{k_1+n \sigma}dk k^2  f_{\chi}(z)\,,
\end{align} 
where the second approximation works for momentum expanding a few $\sigma$, i.e., $n=\mathcal{O}(1)$,  due to the Gaussian peaked distribution function.

We shown in Fig.~\ref{fig:distribution} the comparisons between the  distribution functions obtained from the Mathieu   and Boltzmann equations, respectively. In the top panel, the peak structures of $n_k$ and $f_{\chi}$ are shown for $q=0.001$ and $q=0.1$. In general, $n_k$ and $f_\chi$ are not identical, but there is a large overlap between $n_k$ and $f_\chi$ near the peak $k_1=m_\phi/2$. This  points out that the two distribution functions can reach   approximate consistency between particle production from oscillating $\phi$ in the NPR regime and that from nonrelativistic $\phi$ decay, basically independent of $q$ as long as $q\ll 1$ holds.

The middle panel of Fig.~\ref{fig:distribution} shows the evolution of distribution functions  at the peak $k_1=m_\phi/2$. We see  that $f_{\chi}$ will reach the asymptotic behavior of $n_k$ after the occupation number becomes larger than 1. Such consistency confirms the qualitative interpretation of particle production in the NPR regime. Furthermore, the exponential growth of bosonic particles from oscillating fields can be quantitatively  described by  stimulated decay of nonrelativistic particles.

In the bottom panel of Fig.~\ref{fig:distribution}, we show the evolution of the ratio $n_k/f_\chi$ at the peak $k_1=m_\phi/2$, for $q$ ranging from $10^{-4}$ to $0.1$, confirming that the two functions coincide in the   asymptotic limit $z\gg 1$. For larger $q$, the timescale for reaching the consistency is shorter, since a larger  $q$ will have a larger production rate to create $\chi$ particles,  consequently allowing  faster development of the exponential growth.

We have   shown the consistency between the distribution functions obtained from the Mathieu   and Boltzmann equations only at the peak. As can be inferred from the top panel of Fig.~\ref{fig:distribution}, there are still regions where the two approaches are not identical. However, these regions correspond to the distribution of produced $\chi$  particles away from the peak,  and hence exhibit  suppressed occupation numbers. When integrating over the full momentum space, we may interpret  these non-identical regions as the theoretical errors in predicting   particle production from the Boltzmann equation.

\begin{figure}[t]
	\centering
	\includegraphics[scale=0.55]{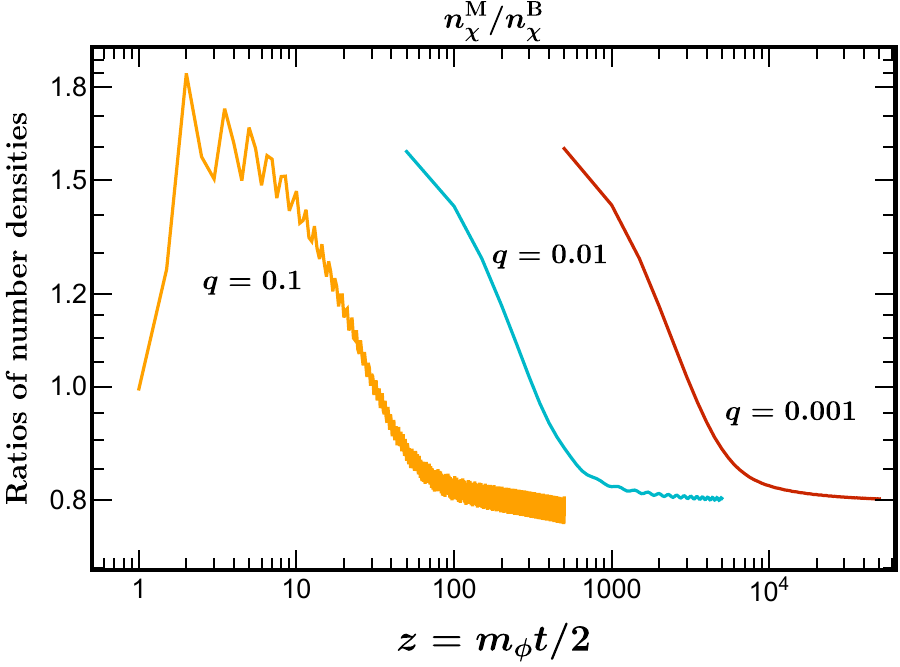} 
	\caption{\label{fig:nz} The time evolution of the ratios $n_\chi^{\rm M}/ n_{\chi}^{\rm B}$, where $n_\chi^{\rm M}$ is given by Eq.~\eqref{eq:nt-def} and  $n_{\chi}^{\rm B}$ is from Eq.~\eqref{eq:nchi-B}.
	}
\end{figure}

 In Fig.~\ref{fig:nz}, we show the discrepancy  via the evolution of the ratios $n_{\chi}^{\rm M}/n_{\chi}^{\rm B}$, where $n_{\chi}^{\rm M}$ denotes the $\chi$ particle number density obtained from the Mathieu equation given in Eq.~\eqref{eq:nt-def}, and $n_{\chi}^{\rm B}$ is obtained in the Boltzmann equation based on Eq.~\eqref{eq:nchi-B}.  We see that  after the ratio $n_k/f_{\chi}$  reaches 1, the ratio $n_{\chi}^{\rm M}/n_{\chi}^{\rm B}$  will   reach  a constant at 0.8, basically independent of $q$. It implies that there is a $20\%$ relative deviation of the particle number density  between the Mathieu and Boltzmann approaches. Nevertheless, such a  deviation, which is also the typical scale between using the Boltzmann and Bose-Einstein distributions,  is usually durable when determining    physical  impacts  on observables.

\section{Validity and application}	\label{sec:disc}
The discussions thus far did not take into account effects from cosmic expansion, backreaction, and   interactions between  the produced $\chi$ particles and a potential thermal plasma. In fact,  it has been widely recognized that these effects can drastically change the exponential growth of the particle distribution function and the number density~\cite{Kasuya:1996np,Kofman:1997yn,Alonso-Alvarez:2019ssa,Carenza:2019vzg,Jaeckel:2021xyo}. In particular, the NPR regime is quite sensitive to momentum redshift and any scattering process that would drive the generated $\chi$ momentum away from the bandwidth $\sigma=q m_\phi$, such that   successful accumulation  of produced particles cannot be prepared to trigger the late-time exponential growth after several periods of oscillations.  Once this situation occurs, we would expect that   NPR fails to generate abundant particles and  the associated observational consequences would hardly arise. This case renders   applications of  the NPR theory limited and   tests of the NPR theory challenging. In addition,  it becomes  less meaningful to discuss the consistency between the NPR and    Boltzmann approaches. 

Therefore, applying the Boltzmann equation to study the exponential growth of particles in an oscillating field background is beneficial for scenarios where   cosmic expansion, backreaction, and   scattering do not prevent the explosive particle production. In the following, we discuss these potential scenarios. 

Cosmic expansion leads to momentum redshift after particles are produced. The timescale for redshifting the momentum spread out of the bandwidth can be estimated via the conserved comoving momentum $k_i a(t_i)=k_f a(t_f)$, with the Hubble parameter $H=a^{-1}da/dt$. Let us focus on the radiation-dominated Universe, where $a(t)\propto \sqrt{t}$ and $H=1/(2t)$. Then the timescale of momentum redshift from the one boundary to the other boundary (see Eq.~\eqref{eq:LOwidth})  yields
\begin{align}
	\Delta t=t_f-t_i=\frac{\sigma}{H_{\rm osc} k}\,,
\end{align}
where we have denoted the Hubble scale as $H_{\rm osc}$ at the initial time $t_i$ or during the   scalar oscillation epoch.  
If this timescale is much shorter than that of the exponential development $\tau_{\rm NPR}\equiv (qm_\phi/2)^{-1}$ (see Eq.~\eqref{eq:fchi-z}), then we should expect no abundant particle production from  NPR. Therefore, for working NPR, we have the condition
\begin{align}\label{eq:redshift}
q^2\gg \frac{H_{\rm osc}}{m_\phi}\,.
\end{align}
Clearly, for $q\ll 1$, a large scale hierarchy $H_{\rm osc}\ll m_\phi$ should be met once the oscillation of $\phi$ begins.

The oscillation time for $\phi$ is determined by the equation of motion
\begin{align}\label{eq:phi-EoM}
	\frac{d^2 \phi}{dt^2} +3H \frac{d\phi}{dt} + \frac{dV_\phi}{d\phi}=0\,.
\end{align}
For scalar potential $V_\phi$ with suppressed higher-order $\phi$ terms, we can approximate  $V_\phi\approx m_\phi^2 \phi^2/2$ so that  $dV_\phi/d\phi=m_\phi^2 \phi$. This situation occurs in  standard QCD axion theories~\cite{Marsh:2015xka,DiLuzio:2020wdo} and in  ultralight dark matter production  via the conventional misalignment mechanism~\cite{Abbott:1982af,Dine:1982ah,Preskill:1982cy}, where the oscillation time is determined by
\begin{align}\label{eq:osc}
m_\phi=3H(T_{\rm osc})\,,
\end{align}
with $T_{\rm osc}$ denoting  the cosmic temperature when the oscillation begins.
It immediately suggests that it is not easy for these scenarios to feature explosive particle production in the NPR regime, since a sufficiently large $q$ is required to overcome the condition~\eqref{eq:redshift} right after the oscillation begins. 

In several realistic applications, cosmic expansion may  not be  the dominant effect in changing the momentum of the produced particles. In particular, when the produced particles are photons, the thermalization timescale in the radiation-dominated Universe is much shorter than the Hubble expansion. The  scattering effect may be estimated by a frequency-dependent medium-induced  mass, or the  thermal mass, which also prevents the development of the  exponential growth~\cite{Abbott:1982af,Alonso-Alvarez:2019ssa}.  Note that, inserting the thermal mass into the Mathieu equation is not well justified~\cite{Lee:1999ae,Yokoyama:2004pf,Yokoyama:2005dv}, since the timescale for developing such a thermal mass is not instantaneous, depending on the interaction rate with the thermal plasma. Nevertheless, for scattering effects that   drive the created momentum away from the bandwidth at a speed faster than  Hubble redshift,  the condition~\eqref{eq:redshift} would be strengthened:
\begin{align}\label{eq:scat}
	q^2\gg \xi \frac{H_{\rm osc}}{m_\phi}\,,
\end{align}
with $\xi\gg1$ depending  on the relative size of the Hubble expansion rate and the scattering rate.  Again in this case,  a large scale hierarchy $H_{\rm osc}\ll m_\phi$ is required in the NPR regime when the oscillation of $\phi$ begins.

The mentioned scale hierarchy can be achieved by scenarios of delayed oscillations. In particular, when $\phi$ has a large initial velocity, the second term in Eq.~\eqref{eq:phi-EoM} cannot be neglected even in the regime $H\ll m_\phi$, such that the oscillation time is not determined by Eq.~\eqref{eq:osc}. This situation occurs in axion production from the  kinetic~\cite{Co:2019jts,Co:2019wyp} or trapped~\cite{DiLuzio:2021gos,DiLuzio:2021pxd} misalignment mechanism, and dark matter production from axion level crossing~\cite{Daido:2015cba,Daido:2015bva}.   

Observational consequences are likely to arise if the  oscillation starts after MeV temperatures, where NPR may leave imprints on sensitive  observables inherited from e.g., neutrino decoupling, BBN, CMB spectral distortions and CMB anisotropies. In this regime, we have 
\begin{align}
\frac{H}{m_\phi}\lesssim 4.4\times \left(\frac{10^{-16}~\text{eV}}{m_\phi}\right)\,,
\end{align}
implying  that an eV-scale mass of $\phi$ can readily satisfy the conditions~\eqref{eq:redshift} and~\eqref{eq:scat} with $q\ll 1$ and hence feature explosive particle production. It is worthwhile to emphasize that for the above mentioned observables, the Boltzmann equation has been well established to make precise predictions. Therefore, applying the Boltzmann approach to the NPR mechanism merits easier evaluation for  the impacts of  explosive particle production on observables.

Once  the conditions~\eqref{eq:redshift} and~\eqref{eq:scat} are satisfied, it is easier to prepare the exponential growth  under several periods of oscillations. Nevertheless, when the generated energy density becomes comparable with the background field,  the backreaction to the $\phi$ field cannot be neglected, which drives  Eq.~\eqref{eq:phi-EoM} into a more involved nonlinear regime~\cite{Khlebnikov:1996mc,Khlebnikov:1996zt}. The backreaction effect is often associated with   the stability of $\phi$, particularly when  $\phi$ is identified with the dark matter candidate. When to consider the backreaction effect also depends on the products and on when   NPR occurs. 

If the products are photons or other relativistic bosonic species such as majorons~\cite{Gelmini:1980re}, the generated energy density should not exceed the   energy budget of the Universe during which primordial observables are formed from   BBN and CMB. In these regimes,  the  effective neutrino number $N_{\rm eff}$ plays  a key observable~\cite{Pitrou:2018cgg,Planck:2018vyg}. Effectively, one can  constrain the generated energy density by requiring 
\begin{align}\label{eq:Rchig}
	R_{\chi\gamma}(z)\equiv \frac{\rho_\chi}{\rho_\gamma}\lesssim \mathcal{O}(0.1)\,,
\end{align}
 where $\rho_\gamma\approx 0.66 T^4$ is the photon energy density, and $\rho_\chi$ can be estimated via the Boltzmann approach
\begin{align}\label{eq:rhochi-B}
	\rho^{\rm B}_{\chi}(z)=\int \frac{d^3 k}{(2\pi)^2}k f_{\chi}(z)\,.
\end{align} 
For relativistic particle production, the condition~\eqref{eq:Rchig} indicates a modification to $N_{\rm eff}$
\begin{align}
\Delta N_{\rm eff}=\frac{8}{7}\left(\frac{11}{4}\right)^{4/3} \left(\frac{\rho_\chi}{\rho_\gamma}\right)\lesssim \mathcal{O}(0.1)\,,
\end{align}
at the order that satisfies current bounds~\cite{Planck:2018vyg} and can be probed in upcoming CMB experiments.

\begin{figure}[t]
	\centering
	\includegraphics[scale=0.55]{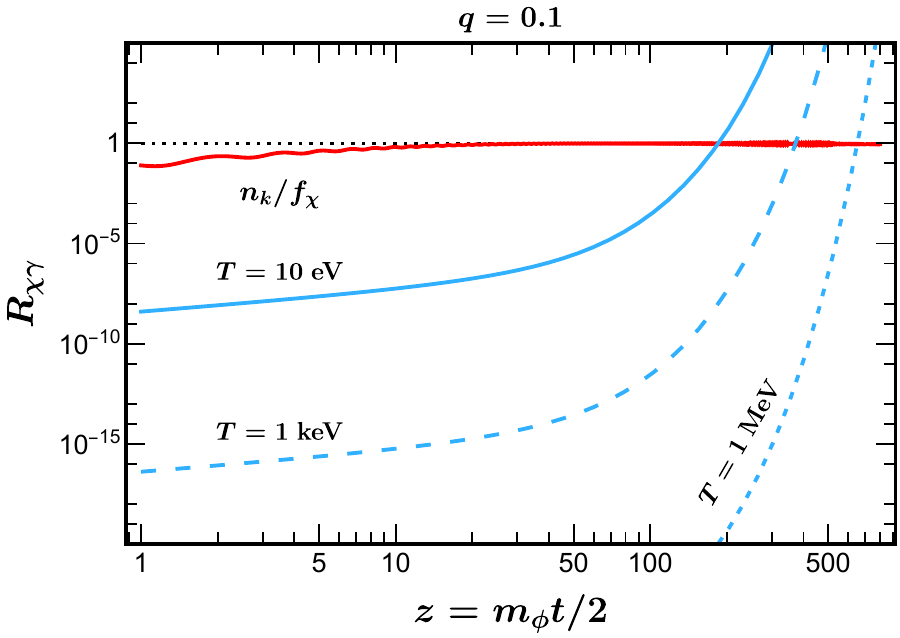} 
	\caption{\label{fig:Rchigam} The evolution of $\chi$ energy density normalized to the photon energy density in different  radiation-dominated epochs, $T=10$~eV, $T=1$~keV, and $T=1$~MeV. It serves to justify the Boltzmann approach with potential observational consequences but without significant backreaction effects. 
	}
\end{figure}

Then one should check if the Boltzmann approach can be well approximately applied  for the whole interested timescale under the condition~\eqref{eq:Rchig}. To see this, we show in Fig.~\ref{fig:Rchigam} the evolution of the ratio $R_{\chi\gamma}$ at different temperatures. We see that while the distribution function $f_\chi$ obtained from the  Boltzmann approach is not exactly equivalent to $n_k$ obtained in    NPR during the  initial periods of oscillations, the generated energy density is initially  too small to cause significant observational consequences. After $\rho_\chi$ becomes larger and approaches  the magnitude of the photon energy density, the Boltzmann approach has already provided a consistent description of the distribution function predicted in   NPR. Therefore, for observational consequences predicted from   NPR at temperatures below a few MeV,  it suffices to use the Boltzmann approach in describing the explosive particle production.  Note that we have fixed $q=0.1$ in Fig.~\ref{fig:Rchigam} for illustration purpose, but  the conclusion is also valid for   $q<0.1$.

Finally, let us comment on  another pattern that satisfies the condition~\eqref{eq:redshift}. This occurs if $q$ is large, $q\gg 1$,  corresponding to the broad parametric resonance. Particle production in this regime also features exponential growth, though in a quite different way~\cite{Fujisaki:1995dy,Fujisaki:1995ua,Khlebnikov:1996wr,Kofman:1997yn,Dufaux:2006ee}, but the resonance width cannot be determined perturbatively. Near the maximum of the field amplitude in a single oscillation, $\phi(t)=\bar \phi$, the frequency $\omega_k(t)$ becomes  purely imaginary since $k^2< \mu \bar \phi$. In this regime, the calculation given in Eq.~\eqref{eq:LOwidth} fails to determine the momentum spread. It  becomes problematic to calculate particle collision rates embedded in the Boltzmann equation. Nevertheless, near the minimum of   the field amplitude where $k^2\gg \mu \phi(t)$, the frequency $\omega_k(t)$   becomes real even for $q\gg 1$. In this regime, particle collision may work to describe the interaction between $\phi$ and $\chi$, and in this case, the higher-order annihilation channels $j\phi\to 2\chi$ become more important than   the decay mode, requiring proper summation over $j$~\cite{Li:2024bbe}.

\section{Conclusion}	\label{sec:con}
We have demonstrated that by simulating the momentum spread with a Gaussian  distribution,  the Boltzmann equation can well describe the exponential growth of the distribution function in bosonic particle production, which is consistent with the prediction of the NPR mechanism in the asymptotic regime. When the particle number density is concerned, the prediction from the Boltzmann approach  only shows a relative discrepancy at $20\%$, rather than orders of magnitude caused by  the mismatch of the exponential growth rate or a different construction of the distribution function. 

We have also discussed the potential regimes where the NPR mechanism may induce observational consequences at cosmic temperatures below  MeV. Under this circumstance, the cosmic expansion, backreaction and scattering with the thermal plasma can be neglected such that explosive particle production is not prevented.    The Boltzmann equation then provides a simple and analytic approach to study the evolution of  distribution functions from oscillating scalar decay, where late-time thermalization processes can be incorporated in a straightforward way.

\section*{Acknowledgements}
The author would like to thank Shinya Kanemura, Kunio Kaneta, Kodai Sakurai, and Ryosuke Sato for helpful discussions and feedback. This work is supported by JSPS Grant-in-Aid for JSPS Research Fellows No. 24KF0060.



\bibliographystyle{JHEP}
\bibliography{Refs}

\end{document}